\documentclass[12pt, fleqn]{article}
\usepackage[cp1251]{inputenc}
\usepackage{latexsym,amsfonts,amssymb}
\usepackage{graphicx}

\usepackage{amsbsy}
\usepackage{amsmath}
\usepackage{epsf}

\newtheorem{theo}{Theorem}
\newcommand{\bt}{\begin{theo}}
\newcommand{\et}{\end{theo}}
\newcommand{\bd}{\begin{displaymath}}
\newcommand{\ed}{\end{displaymath}}

\newcommand{\lf}{\left}
\newcommand{\rg}{\right}

\newcommand{\be} {\begin{equation}}
\newcommand{\ee} {\end{equation}}
\newcommand{\ba} {\begin{array}}
\newcommand{\ea} {\end{array}}

\newcommand{\p} {\partial}

\newcommand{\vp} {\varphi}

\newcommand{\al} {\alpha}
\newcommand{\lbd} {\lambda}

\begin{document}
\normalsize
 \begin{center}
 {\Large \bf New conditional symmetries
 and exact solutions \\ of
 reaction-diffusion systems with power diffusivities }\footnote{\small to appear in J.Phys.A: Math.Theor.}\\
 \medskip
 
{\bf Roman Cherniha$^{\dag,\ddag}$}
 {\bf and  Oleksii Pliukhin$^\dag$}
 \\
{\it  $^\dag$~Institute of Mathematics, Ukrainian National Academy
of Sciences,\\
 Tereshchenkivs'ka Street 3, Kyiv 01601, Ukraine}\\
 {\it $^\ddag$~Lesya Ukrayinka Volyn' National University,\\
13, Volia Avenue,   Lutsk 43025, Ukraine}\\
 \medskip
 E-mail: cherniha@imath.kiev.ua and pliukhin@imath.kiev.ua
\end{center}

\begin{abstract}
 A wide range of new Q-conditional symmetries for
reaction-diffusion systems with power diffusivities is constructed.
The relevant non-Lie ans\"atze  to reduce the 
reaction-diffusion systems to ODE systems  and  examples of exact solutions  are obtained.
Relation of the solutions obtained with  the development of spatially inhomogeneous  structures
is discussed.
\medskip\\
PACS numbers: 02.20.-a, 02.30.Jr, 05.45.-a
\end{abstract}

\begin{center}
{\bf 1.Introduction.}
\end{center}

In 1952 A.C. Turing published the remarkable paper \cite{turing}, in which  a revolutionary
idea about mechanism of morphogenesis (the development of structures in an organism during the life)
has been proposed.
Roughly speaking,  his idea says that the diffusion process can interact with the chemical reaction
process in such a way that this can stimulate   the development and growth of different forms
and structures  in an organism. Moreover this differentiation is often impossible without diffusion,
i.e., only one may destabilize the spatially homogeneous  structures. This effect is known as the Turing
instability and mathematically leads to systems of reaction-diffusion equations
with nonlinearities  of the  special form \cite{mur2, britton}.
 Since 1952 the reaction-diffusion systems have been extensively
studied by means of different mathematical methods, including group-theoretical methods.
The main attention was paid to investigation of
 the two-component RD systems of the form
 \be\label{1*}\ba{l}
 U_t=[D^1(U)U_x]_x+F(U,V),\\
V_t=[D^2(V)V_x]_x+ G(U,V)
 \ea\ee
where
  $U= U(t,x)$ and $V= V(t,x)$ are two  unknown functions representing the  densities of cells and chemicals,
$F(U,V)$ and $G(U,V)$ are two given functions describing interaction between them and environment,
 the functions $D^1(U)$ and $D^2(V)$ are the relevant diffusivities (usually they are assumed
 to be constants or power functions), and the subscripts $t$ and $x$ denote
differentiation with respect to these variables. At the present time
one can claim that all possible  Lie symmetries of (\ref{1*}) with
the constant diffusivities are completely described in
\cite{ch-king,ch-king2,niki-05}, while in
\cite{ibrag-94,ch-king4,ch-king5}    it has been done for  the non-constant
diffusivities.

The problem
 of construction of conditional symmetries for (1)  is  still not
  solved even in the case of $Q$-conditional symmetries (non-classical symmetries).
  Moreover, to  our best knowledge, there are no any papers
  devoted to the search of conditional symmetries  of the  RD system (\ref{1*}).
 Here, we present some recent results in this direction for the first time.
 It should be noted that there are many papers  devoted to the construction of  such symmetries
 for the  {\it scalar} non-linear reaction-diffusion (RD) equations of the form
 \cite{fss, se-90,  nucci1, cla, a-b-h, a-h, pucci-sac2000}
 \be\label{1**}
 U_t=\left[D(U)U_x\right]_x + F(U).
\ee
and (\ref{1**}) with the  convective term   $B(U)U_x$ (here $B(U), D(U)$ and $F(U)$ are  arbitrary smooth
functions) \cite{ ch-se-98,che-2006,ch-pliu-2006}.

It is well-known that conditional symmetries can be applied for
finding exact solutions of the relevant equations, which are not
obtainable by the classical Lie method. Moreover the solutions
obtained in such a way may have   a  physical or biological
interpretation (see, e.g.,  examples in  \cite{
che-2006,ch-pliu-2006,dix-cla, broad-2004}). In this paper we will demonstrate
this using the  $Q$-conditional symmetries  of non-linear RD
systems of the form
 (\ref{1*}).

  The paper is organized as follows.
  In the second section, we present  new  $Q$-conditional symmetries of the RD system
\be\label{1}\ba{l}
U_t=(U^kU_x)_x+F(U,V),\\
V_t=(V^lV_x)_x+ G(U,V),\ea\ee
which is the most important subcase of (\ref{1*}) from the applicability point of view \cite{mur2, okubo}.
The determining equations for constructing $Q$-conditional symmetries of
system (\ref{1}) are derived and   the main theorem presenting these symmetries in explicit form
is proved.

In the third  section, the $Q$-conditional symmetries obtained are applied for  reducing the corresponding RD systems
to the systems of ordinary differential equations (ODE). A nonlinear  RD system  with the specified functions 
 $F(U,V)$ and $G(U,V)$ is considered in details. We show that this system with the correctly-specified
 coefficients may lead to the  Turing  instability. Finally, we present some conclusions.

\begin{center}
{\textbf{2. Main result.}} \end{center}

It is well-known that the main difficulty arising in search of $Q$-conditional symmetries
is to solve so called system of determining equations, which is nonlinear and overdetermined.
Roughly speaking, one arrives on contradiction trying to construct   $Q$-conditional symmetries
of a given nonlinear PDE (system of PDEs) because there is the need to solve new nonlinear
PDE system, which usually is much more cumbersome. This problem arises even in the case
of linear PDE and it was the reason why G.Bluman and J.Cole in their pioneering work \cite{bl-c}
were unable to describe all  $Q$-conditional symmetries in explicit form even for the linear heat
equation. In the recent paper \cite{ch-pliu-2006},  a short historical review is devoted
to this problem and different kinds of non-Lie symmetry.

Here, we construct the  system of determining equations to find $Q$-conditional symmetry operators
of the form
\begin{equation}
\label{10}
 {Q}= \ \partial_t+\xi(t,x,U,V)\partial_x+\eta^1(t,x,U,V)\partial_U+\eta^2(t,x,U,V)\partial_V,
\end{equation}
where $\xi$, $\eta^1$ and $\eta^2$ are unknown functions, for the RD
system  (\ref{1}) and solve one under the relevant restrictions.
Note that  we search for purely
 conditional symmetry operators, which  cannot be reduced
 to Lie symmetry operators. We also remind the reader that each $Q$-conditional symmetry operator
 is determined up to equivalent representations generated by multiplying
   (\ref{10})   on the arbitrary smooth function $M(t,x,U,V)$.
   The problem of finding $Q$-conditional symmetry operators
of the form
\begin{equation}
\label{10*}
 {Q}= \ \xi(t,x,U,V)\partial_x+\eta^1(t,x,U,V)\partial_U+\eta^2(t,x,U,V)\partial_V,
\end{equation}
is another challenge and  will be treated elsewhere.

First of the all, we remind the reader the notion of the conditional symmetry following
 the book \cite [Section 5.7]{fss} (note one  was formulated for the first time  in \cite{Fush88}).

\noindent {\bf Definition.}  A PDE 
 of the form
 \be\label{2:gl:c1} S\left(x_1,...,x_r,U,\underset{1}{U},...
\underset{p}{U}\right) =0 \ee
(here $U=U(x_1,...,x_r)$ and $\underset{k}{U}$ is the
totality of $k^{\rm th}$-order derivatives) is {\it conditionally
invariant} under the operator
\be\label{2:gl:Q}
Q = \xi^1 (x_1,...,x_r, U)\partial_{x_1} +...+\xi^r (x_1,...,x_r, U)\partial_{x_r}+ \eta(x_1,...,x_r,U)\partial_U,
\ee
where $\eta$ and $\xi^a$ with $a=1,\ldots,r$ are smooth functions,
if it is invariant (in Lie's sense) under this operator only
together with an additional condition of the form
\be
\label{2:gl:c2} S_Q\left(x_1,...,x_r,U,\underset{1}{U},\ldots
\underset{q}{U}\right) =0,
\ee
that is, the overdetermined system of
equations (\ref{2:gl:c1}) and (\ref{2:gl:c2}) is invariant under a Lie group
generated by the operator $Q$. If the additional condition (\ref{2:gl:c2})
coincides with the equation $Q(U)=0$, i.e., 
\[
\xi^1 (x_1,...,x_r, U)U_{x_1}+...+ \xi^1 (x_1,...,x_r, U)U_{x_r}= \eta(x_1,...,x_r,U)
\]
then PDE (\ref{2:gl:c1}) is {\it Q-conditionally
invariant} under the operator (\ref{2:gl:Q}).

Obviously, this definition admits a direct generalization on systems of two PDEs.
Of course,  if the given system is not in involution then some difficulties can occur,
however, system  (\ref{1}) has the involution form. Let us show how the determining equations
to find   $Q$-conditional symmetry operator (\ref{10}) are obtained.

It turns out that one can apply the local substitution
 \be\label{2}\ba{l} u=U^{k+1},\ k\ne-1,\\v=V^{l+1},\
l\ne-1.\ea\ee to simplify the further computations. Of course, the
case $(k+1)(l+1)=0$ is special and needs of a separate investigation.
Substitution  (\ref{2}) reduces system (\ref{1}) and operator
(\ref{10}) to the form
\be\label{3}\ba{l}u_{xx}=u^mu_t+C^1(u,v),\\v_{xx}=v^nv_t+C^2(u,v),\ea\ee
and
\be\label{5}Q=\p_t+\xi(t,x,u,v)\p_x+\eta^1(t,x,u,v)\p_u+\eta^2(t,x,u,v)\p_v,\ee
where  $m=-\frac{k}{k+1}\ne-1,\ n=-\frac{p}{p+1}\ne-1,\
C^1(u,v)=-(k+1)F\lf(u^{\frac{1}{k+1}},v^{\frac{1}{l+1}}\rg),\\ \
C^2(u,v)=-(p+1)G\lf(u^{\frac{1}{k+1}},v^{\frac{1}{l+1}}\rg), \p_u=
\frac{1}{k+1}U^{-k}\p_U, \p_v= \frac{1}{l+1}V^{-l}\p_V$. 


Let us apply now the second prolongation of the operator $Q$
\[ \ba{l} \medskip
\mbox{\raisebox{-1.1ex}{$\stackrel{\displaystyle Q}{\scriptstyle 11}$}}
=Q+\rho^1_t\frac{\p}{\p u_t}+\rho^2_t\frac{\p}{\p v_t}+\rho^1_x\frac{\p}{\p u_x}+\rho^2_x\frac{\p}{\p v_x} \\
\qquad +\sigma^1_{tx}\frac{\p}{\p u_{tx}}+\sigma^2_{tx}\frac{\p}{\p v_{tx}}
+\sigma^1_{tt}\frac{\p}{\p u_{tt}}+\sigma^2_{tt}\frac{\p}{\p v_{tt}}
+\sigma^1_{xx}\frac{\p}{\p u_{xx}}+\sigma^2_{xx}\frac{\p}{\p v_{xx}}, \ea \]
where coefficients $\rho^k$ and $\sigma^k$ with the relevant indices are calculated by the 
known formulas (see, e.g., \cite{fss, ol}), to the each equations 
of (\ref{3}). Since system  
(\ref{3}) should be  considered as a manifold 
in the space of independent variables
\[
t, \, x, \,u, \,v,\, u_t, \,v_t, \,u_{x},\, v_x, \,u_{xt}, \,v_{xt}, \,u_{tt}, \,v_{tt},
\,u_{xx}, \,v_{xx},\]
we arrive at the invariance condition
 \be\label{4}\ba{l} m \eta^1 u^{m-1}
u_t+\eta^1 C^1_u+\eta^2 C^1_v+\rho^1_t u^m =\sigma^1_{xx},\\
n \eta^2v^{n-1} v_t+\eta^2 C^2_v+\eta^1 C^2_u+\rho^2_t
v^n=\sigma^2_{xx}.\ea\ee 
To obtain the determining equation for 
$Q$-conditional symmetry operator (\ref{5}), one needs to take into
account not only system (\ref{3}) (it will lead only to 
the determining equation for Lie symmetry operators !) 
but two additional conditions 
\be\label{10-1}u_t+ \xi u_x=\eta^1,\ v_t+\xi v_x=\eta^2 \ee 
generated by operator (\ref{5}).
Thus, inserting into (\ref{4}) the explicit expression for $\rho^k$ and $\sigma^k$, namely:
   \be\label{10-2}\ba{l}\medskip
\rho^1_t=\eta^1_t+\eta^1_u u_t +\eta^1_vv_t-u_x(\xi_t+\xi_uu_t
+\xi_vv_t),\\ \medskip\rho^2_t=\eta^2_t+\eta^2_u u_t
+\eta^2_vv_t-v_x(\xi_t+\xi_uu_t +\xi_vv_t),\\
\sigma^1_{xx}=\eta^1_{xx}+2\eta^1_{xu}u_x+2\eta^1_{xv}v_x+\eta^1_{uu}(u_x)^2+
\eta^1_{vv}(v_x)^2+2\eta^1_{uv}u_xv_x+\eta^1_uu_{xx}+\eta^1_vv_{xx}-\\
\qquad\quad-u_x(\xi_{xx}+2\xi_{xu}u_x+2\xi_{xv}v_x+\xi_{uu}(u_x)^2+
\xi_{vv}(v_x)^2+2\xi_{u v}u_xv_x+\xi_u u_{xx}+\xi_vv_{xx})-\\
\medskip \qquad\quad-2u_{xx}(\xi_x+\xi_u
u_x+\xi_vv_x),\\
\sigma^2_{xx}=\eta^2_{xx}+2\eta^2_{xu}u_x+2\eta^2_{xv}v_x+\eta^2_{uu}(u_x)^2+
\eta^2_{vv}(v_x)^2+2\eta^2_{uv}u_xv_x+\eta^2_uu_{xx}+\eta^2_vv_{xx}-\\
\qquad\quad-u_x(\xi_{xx}+2\xi_{xu}u_x+2\xi_{xv}v_x+\xi_{uu}(u_x)^2+
\xi_{vv}(v_x)^2+2\xi_{u v}u_xv_x+\xi_u u_{xx}+\xi_vv_{xx})\\
\qquad\quad-2v_{xx}(\xi_x+\xi_u u_x+\xi_v v_x)\ea\ee
and excluding four derivatives $u_t, v_t, u_{xx},v_{xx}$ using
(\ref{3}) and (\ref{10-1}), one arrives at the  cumbersome expressions

 \be\label{10-3}\ba{l}
u^m\lf(\eta^1_t+\eta^1_u (\eta^1-\xi u_x) +\eta^1_v(\eta^2-\xi
v_x)-u_x\Bigr(\xi_t+\xi_u(\eta^1-\xi u_x) +\xi_v(\eta^2-\xi
v_x)\Bigr)\rg)+\\+m\eta^1u^{m-1} (\eta^1-\xi u_x)+\eta^1
C^1_u+\eta^2 C^1_v=
\eta^1_{xx}+2\eta^1_{xu}u_x+2\eta^1_{xv}v_x+\eta^1_{uu}(u_x)^2+
\eta^1_{vv}(v_x)^2+\\+2\eta^1_{uv}u_xv_x-u_x(\xi_{xx}+2\xi_{xu}u_x+2\xi_{xv}v_x+\xi_{uu}(u_x)^2+
\xi_{vv}(v_x)^2+2\xi_{u v}u_xv_x)+\\ \medskip+\Bigr((\eta^1-\xi
u_x)u^m+C^1\Bigr)(\eta^1_u-2\xi_x-3\xi_uu_x-2\xi_vv_x)+\Bigr((\eta^2-\xi
v_x)v^n+C^2\Bigr)(\eta^1_v-\xi_vu_x),\\
v^n\lf(\eta^2_t+\eta^2_u (\eta^1-\xi u_x) +\eta^2_v(\eta^2-\xi
v_x)-v_x\Bigr(\xi_t+\xi_u(\eta^1-\xi u_x) +\xi_v(\eta^2-\xi
v_x)\Bigr)\rg)+\\+n\eta^2v^{n-1} (\eta^2-\xi v_x)+\eta^1
C^2_u+\eta^2 C^2_v=
\eta^2_{xx}+2\eta^2_{xu}u_x+2\eta^2_{xv}v_x+\eta^2_{uu}(u_x)^2+
\eta^2_{vv}(v_x)^2+\\+2\eta^2_{uv}u_xv_x-v_x(\xi_{xx}+2\xi_{xu}u_x+2\xi_{xv}v_x+\xi_{uu}(u_x)^2+
\xi_{vv}(v_x)^2+2\xi_{u v}u_xv_x)+\\+\Bigr((\eta^2-\xi
v_x)v^n+C^2\Bigr)(\eta^2_u-2\xi_x-3\xi_vv_x-2\xi_uu_x)+\Bigr((\eta^1-\xi
u_x)u^m+C^1\Bigr)(\eta^2_u-\xi_uv_x). \ea\ee
 Finally, we take into
account that the unknown functions $\eta^1, \eta^2$ and $\xi$ don't
depend on the derivatives $u_x$ and $v_x$ and therefore we split two
expressions arising in (\ref{10-3}) on
 $(u_x)^3,\ u_x(v_x)^2,\ v_x
(u_x)^2,\ u_xv_x,\ (u_x)^2,\ (v_x)^2,\ v_x,\ u_x $ and
$(v_x)^3,\ v_x(u_x)^2,\ u_x (v_x)^2,\ u_xv_x,\ (u_x)^2,\ (v_x)^2,\
v_x,\ u_x, $ respectively.
Thus,  we arrive at the nonlinear  system
of determining equations
\be\label{6}\ba{l} 1)\ \medskip \xi_{uu}=\xi_{vv}=\xi_{uv}=0,
\\2)\ \medskip \eta^1_{vv}=0,
\\3)\ \medskip \eta^2_{uu}=0,
\\4)\ \medskip 2\xi\xi_u u^m+\eta^1_{uu}-2\xi_{xu}=0,
\\5)\ \medskip 2\xi\xi_v v^n+\eta^2_{vv}-2\xi_{xv}=0,
\\6)\ \medskip \xi\xi_v(u^m+v^n)+2\eta^1_{uv}-2\xi_{xv}=0,
\\7)\ \medskip \xi\xi_u(u^m+v^n)+2\eta^2_{uv}-2\xi_{xu}=0,
\\8)\ \medskip \xi\eta^1_v(u^m-v^n)+2\eta^1_{xv}-2\xi_v C^1-2\xi_v\eta^1u^m=0,
\\9)\ \medskip \xi\eta^2_u(v^n-u^m)+2\eta^2_{xu}-2\xi_u C^2-2\xi_u\eta^2v^n=0,
\\10)\ \medskip -m\xi\eta^1u^{m-1}+(2\xi_u\eta^1-\xi_t-\xi_v\eta^2-2\xi\xi_x)u^m+
\\ \medskip +\xi_v\eta^2v^n+3\xi_uC^1+\xi_vC^2-2\eta^1_{xu}+\xi_{xx}=0,
\\11)\ \medskip -n\xi\eta^2v^{n-1}+(2\xi_v\eta^2-\xi_t-\xi_u\eta^1-2\xi\xi_x)v^n+
\\ \medskip +\xi_u\eta^1u^m+3\xi_vC^2+\xi_uC^1-2\eta^2_{xv}+\xi_{xx}=0,
\\12)\ \medskip m(\eta^1)^2u^{m-1}+(\eta^1_t+\eta^2\eta^1_v+2\xi_x\eta^1)u^m-\eta^2\eta^1_vv^n+
\\ \medskip + \eta^1C^1_u+\eta^2C^1_v-\eta^1_uC^1+2\xi_xC^1-\eta^1_vC^2-\eta^1_{xx}=0,
\\13)\ \medskip n(\eta^2)^2v^{n-1}+(\eta^2_t+\eta^1\eta^2_u+2\xi_x\eta^2)v^n-\eta^1\eta^2_uu^m+
\\ \medskip +\eta^1C^2_u+\eta^2C^2_v-\eta^2_uC^1+2\xi_xC^2-\eta^2_vC^2-\eta^2_{xx}=0.
\ea\ee
to find the coefficients of the operator (\ref{5}) and the functions $C^1, C^2.$
Equations 1) from this  system are easily integrated and lead to
\be\label{31}\xi=a(t,x)u+b(t,x)v+c(t,x),\ee
where $a, b$ and $c$ are arbitrary (at the moment) smooth functions.
Substituting (\ref{31}) into  equations 6) and 7) of (\ref{6}) and taking into account
the second and third equations of (\ref{6}), one arrives at the requirement
$a=b=0$. 
 Thus, equations 2)--7) of system (\ref{6}) can be
straightforward integrated  and their general solution takes the
form
\be\label{6*}\ba{l}
\xi=c(t,x),\\
\eta^1=q^1(t)v+r^1(t,x)u+p^1(t,x),\\
\eta^2=q^2(t)u+r^2(t,x)v+p^2(t,x),\ea\ee
where the functions in right-hand-side  are arbitrary those of their arguments.

The remaining equations
8)-13) of system  (\ref{6}) involve the unknown functions $C^1$ and $ C^2$ and are called
the classification equations.
To solve them one should consider three different cases depending on the functions $q^1(t),\ q^2(t)$ and $ \xi(t,x)$
arising in
(\ref{6*}):

a) $q^1(t)=q^2(t)=0,\ \xi(t,x)\ne0;$

b) $q^1(t)=q^2(t)=0,\ \xi(t,x)=0;$

c) $q^1(t)^2+q^2(t)^2\ne0,\ \xi(t,x)=0.$

Note that the fourth possible case
 $q^1(t)^2+q^2(t)^2\ne0,\ \xi(t,x)\not=0$ arises only under  the restriction
 $m=n=0$, which  follows from 8)-9) of system  (\ref{6}).
 It means 
  the case of the RD system (\ref{1}) with
constant diffusivities should be separately considered so that  we assume $m^2+n^2\not=0$ below.

 Solving equations
8)-13) of system  (\ref{6}) in the case (a) is rather simple but very cumbersome.
After the relevant computations ( the program package MATHEMATICA 5.0 was also used)
it has been  established that all the operators obtained
 are nothing else but the Lie symmetry operators (up to the relevant multiplier $M(t,x,U,V)$) found earlier 
in \cite{ch-king4}. In contrary to  (a), the case (c) is the most difficult
 and at the present time we were able to solve equations
8)-13) of system  (\ref{6}) only in particular subcases.

Now we present the theorem giving a complete description of $Q$-conditional symmetry operators
of the form (\ref{10}) in the case (b).

\begin{theo}
 The RD system  (\ref{1}) with $k^2+l^2\ne 0$ and $(k+1)(l+1)\not=0$ is $Q$-conditional invariant under
the operator (\ref{10}) with $\xi=0$ and $\eta^1_V=\eta^2_U=0$ if
and only if
 it  and  the relevant operator (up to equivalent representations generated by multiplying
      on the arbitrary smooth function $M(t,x,U,V)$) have the   forms listed in the table 1
      (here $f$ and $g$ are arbitrary smooth functions of the relevant argument, while $\lambda_j, j=1,2,3,4$ are arbitrary constants).

\end{theo}

\newpage

{\bf Table 1.   $Q$-conditional symmetries of the RD system
(\ref{1})} \footnotesize
\begin{center}
\begin{tabular}{|c|c|c|c|c|}
\hline
no & RD systems of the form (\ref{1*}) &Q-conditional operators& Restrictions    \\
\hline
 &&& \\
1. & $U_t=(U^kU_x)_x+f(U^{k+1}),$ &  $\partial_t+2p(x)V^{1\over2}\partial_V$ & $p_{xx}=(p)^2+\lambda p,$ \\
  &$   V_t=(V^{-\frac{1}{2}}V_x)_x -2 \lambda V^{1\over2}+g(U^{k+1})$& &$  p\ne0$ \\

\hline
 &&& \\

2. & $U_t=(U^kU_x)_x+\lambda_1 U^{-k}+ f(U^{k+1}-\alpha V^{l+1}), $ & $ \partial_t+\lambda_1U^{-k}\partial_U+$ &$\alpha=\frac{\lambda_1(k+1)}{\lambda_2(l+1)},\ \lambda_2\ne0$  \\
  &$  V_t=(V^{l}V_x)_x + \lambda_2 V^{-l}+g(U^{k+1}-\alpha V^{l+1}) $   & $+\lambda_2 V^{-l}\partial_V$ &$
\lbd_1^2+l^2\ne0.$\\
\hline
 &&& \\

3. & $U_t=(U^{-{1\over2}}U_x)_x-2\lambda U^{1\over2}+ f(U^{1\over2}- V^{1\over2}), $ 
& $\partial_t+$ & $p_{xx}=(p)^2+\lambda p, $ \\
  &$ V_t=(V^{-\frac{1}{2}}V_x)_x-2 \lambda V^{1\over2}+g(U^{1\over2}- V^{1\over2}) $  
   & $+2p(x)(U^{1\over2}\partial_U+V^{1\over2}\partial_V)$ & $ p\ne0 $ \\
\hline
 &&& \\

4.&  $ U_t=(U^kU_x)_x+\lambda_1 U^{-k}+ f(\omega), $ &$\partial_t+\lambda_1U^{-k}\partial_U+$ & $\omega=\frac{\exp U^{k+1}}{(V^{l+1}-\lambda_3)^{\frac{\lambda_1(k+1)}{\lbd_2(l+1)}}},\ \lbd_2\ne0,$ \\
 & $V_t=(V^lV_x)_x+(V^{l+1}-\lambda_3)(g(\omega)+\lambda_2 V^{-l}) $
 &$+\lambda_2(V-\lambda_3V^{-l})\partial_V$&$either\ \lbd_1^2+\lbd_3^2\ne0\ or $\\
&$  $& &$  \lbd_3^2+k^2\ne0\ or \lbd_1^2+l^2\ne0$\\ \hline
 &&& \\

5.&  $ U_t=(U^kU_x)_x+( U^{k+1}-\lambda_1)(f(\omega)+\lambda_2
U^{-k})  $ &
$\partial_t+\lambda_2(U-\lambda_1U^{-k})\partial_U+$  & $\omega=\frac{U^{k+1}-\lambda_1}{(V^{l+1}-\lambda_3)^{{\lambda_2(k+1)}\over{\lambda_4(l+1)}}},\ \lambda_2 \lambda_4\ne0,$ \\
 & $ V_t=(V^lV_x)_x+(V^{l+1}-\lambda_3)(g(\omega)+\lambda_4 V^{-l})$ &$+\lambda_4(V-\lambda_3
V^{-l})\partial_V$ & $either\ \lbd_1^2+\lbd_3^2\ne0,\ $  \\
&$  $& &$ or \ \lbd_3^2+k^2\ne0\ or \ \lbd_1^2+l^2\ne0$\\

\hline
\end{tabular}
\end{center}
\normalsize

\textbf{Sketch of the proof of Theorem.}
To prove the theorem one needs to construct  the general solution of
subsystem 8)-13) of system  (\ref{6}) under the assumption $\xi=0$ and $\eta^1_V=\eta^2_U=0$.
Obviously, equations 8) and 9) are automatically satisfied, while those  10) and 11)
are reduced to the form $\eta^1_{xU}=0$ and $\eta^2_{xU}=0,$ respectively, i.e.:
\[ r^1=r^1(t),\quad  r^2=r^2(t).\]
So, the last equations
12) and  13)  take the form
 \be\label{8}\ba{l}(r^1u+p^1)C^1_u+(r^2
v+p^2)C^1_v-r^1 C^1+\\\medskip+(r^1_t+m
(r^1)^2)u^{m+1}+(p^1_t+2 m r^1p^1)u^m+m(p^1)^2u^{m-1}-p^1_{xx}=0,\\
(r^1u+p^1)C^2_u+(r^2 v+p^2)C^2_v-r^2 C^2+\\+(r^2_t+n
(r^2)^2)v^{n+1}+(p^2_t+2 n
r^2p^2)v^n+n(p^2)^2v^{n-1}-p^2_{xx}=0.\ea\ee

System (\ref{8}) consists of two independent first-order linear PDE with respect to the
unknown functions  $C^1(u,v)$ and $C^2(u,v)$ therefore its general solution can be straightforward
constructed, however we should remember that the coefficient in (\ref{8})
are functions on $t$ and $x$. To construct all possible solutions of (\ref{8})
one needs to consider six different cases (up to renaming $u \to v$ and $v \to u$):

1) $r^1=p^1=r^2=p^2=0,$

2) $r^1=p^1=0,\ r^2\ne0,$

3) $r^1=p^1=r^2=0,\ p^2\ne0,$

4) $r^1=0,\ p^1\ne0,\ r^2=0,\ p^2\ne0,$

5) $r^1=0,\ p^1\ne0,\ r^2\ne0,$

6) $r^1\ne0,\ r^2\ne0.$

In the case 1)  operator (\ref{10}) immediately takes the form $Q=\p_t,$
which is, of course, the Lie symmetry operator. The similar situation occurs
in the case 2) since all the  operators obtained are equivalent to
the relevant Lie symmetry operators listed in  \cite{ch-king4}.
The most interesting cases are those 3)--6). 

Consider case 3) in details.
In this case system
 (\ref{8}) takes the form
\be\label{9}\ba{l} p^2C^1_v=0,\\ p^2
C^2_v+p^2_tv^n+n(p^2)^2v^{n-1}-p^2_{xx}=0\ea\ee
and its formal integration  leads to the solution
\be\label{9*}\ba{l}
C^1=f(u)\\
C^2=\int\lf({p^2_{xx}\over
p^2}-{p^2_t\over p^2}v^n-n p^2v^{n-1}\rg)dv + g(u), \ea\ee
where $f$ and $g$ are arbitrary smooth functions.
Since the function $C^2$ doesn't depend on on $t$ and $x$,
three subcases should be separately examined: $n=0$,  $n=1$ and $n\not=0; 1.$
The first subcase immediately gives
$ C^2={{p^2_{xx}-p^2_t}\over p^2}v+g(u),$ so that
\[{{p^2_{xx}-p^2_t}\over p^2}=\lbd, \]
where  $\lbd $ is an arbitrary constant. So, the system
\be\label{21}\ba{l}u_{xx}=u^mu_t+f(u),\\ v_{xx}=v_t+\lbd v+g(u)
\ea\ee 
admits the $Q$-conditional symmetry operator
\be\label{21*}
 Q=\p_t+p^2(t,x)\p_v,\ee
 where $p^2(t,x)$ is the general solution of the linear PDE $ p^2_t=p^2_{xx}-\lbd p^2.$
However, if one now applies substitution (\ref{2}) to
 (\ref{21}) and (\ref{21*}), then the RD system and the operator listed
 in  \cite{ch-king4} (see  case 5 in  table 1) are obtained. So, subcase $n=0$
 doesn't lead to any $Q$-conditional symmetries.

 In the subcase $n=1$ the general solution of (\ref{9}) takes the form
  \[  C^1=f(u), \quad C^2=\lbd v+g(u),\]
  where $\lbd={p^2_{xx}\over p^2}-p^2.$
  So, the system
\be\label{11}\ba{l}u_{xx}=u^mu_t+f(u),\\ v_{xx}= v v_t+\lbd v+g(u)\\
\ea\ee
admits the $Q$-conditional symmetry operator
\be\label{11*}
Q=\p_t+p^2(x)\p_v, \ee
where the function  $p^2(x)$ is the general solution of the nonlinear ODE
\be\label{11**} p^2_{xx}=(p^2)^2+\lbd p^2. \ee
Applying now  substitution (\ref{2}) to
 (\ref{11})--(\ref{11*}) and introducing the relevant notations, one arrives at the system and
 the $Q$-conditional operator listed in case 1 of  table 1.

 Considering the subcase $n\ne0;1,$ we immediately obtain $p^2=\lbd =const$ (see (\ref{9*}))
 and this lead to the system
\be\label{12}\ba{l}u_{xx}=u^mu_t+f(u),\\ v_{xx}= v^n v_t-\lbd v^n+g(u)
\ea\ee
and the operator
\be\label{12*} Q=\p_t+\lbd \p_v. \ee
 Operator (\ref{12*}) is reduced to the form
$Q=\p_t+\frac{1}{l+1}\lbd V^{-l}\p_V$ with $\lbd\not=0$ by using substitution (\ref{2}). 
On the other hand, system (\ref{12}) and operator (\ref{12*}) correspond  to a
particular case at $\lbd_1=\alpha=0$ of those listed in case 2 of table 1.
Thus, case 3) is completely investigated.

The case 4) can be examined in a quite similar way and the system
 \be\label{13}\ba{l}u_{xx}=u^mu_t-\al \lbd u^m+f(u-\al v),\\
v_{xx}=v^n v_t-\lbd v^n+g(u-\al v),\\
Q=\p_t+\lbd(\al\p_u+\p_v),\ea\ee
and the operator
 \be\label{13*}
Q=\p_t+\lbd(\al\p_u+\p_v) \ee
are obtained, where $\al\ne0$ is an arbitrary constant.
It is easily seen that systems and operators
 (\ref{12})--(\ref{13*}) can be united, i.e., the restriction $\al\ne0$ is not essential.
 Applying now  substitution (\ref{2}) to
 (\ref{13})--(\ref{13*}) and introducing the relevant notations, one arrives at the system and
 the $Q$-conditional operator listed in case 2 of  table 1.
 It turns out that the power $m=n=1$ leads to an additional symmetry in this case.
In fact, the system
 \be\label{14}\ba{l} u_{xx}=u u_t+\lbd u+f(u-v),\\
v_{xx}=v v_t+\lbd v+g(u-v) \ea\ee
is conditionally invariant with respect to the operator
\be\label{14*} Q=\p_t+p^2(x)(\p_u+\p_v),\ee
where $p^2(x)$ is the general solution of the nonlinear ODE (\ref{11**}).
Formulas (\ref{14})--(\ref{14*}) together with the substitution  (\ref{2})
generate the RD  system and the operator listed in case 3 of  table 1.
Those listed in cases 4 and 5 of the table can be similarly obtained by examination
of the cases 5) $r^1=0,\ p^1\ne0,\ r^2\ne0$ and  6) $r^1\ne0,\ r^2\ne0.$

Finally, we note that all operators arising in table 1 are not   Lie symmetry operators
because any Lie symmetry operator of the RD system (\ref{1}) must be linear on $U$ and $V$
\cite{ch-king4}. 

 {\it The sketch of the proof is now completed.}

  {\textbf{ Remark 1.}} Restrictions on the coefficients $\lambda_k$ arising in the last column of table 1
  guarantee that the relevant operators do not coincide with   Lie symmetries. Of course,
  those operators is still $Q$-conditional symmetry operator if some lambda-s vanish and then
  they are equivalent to the relevant Lie symmetry operators obtained in   \cite{ch-king4}.

\newpage

\begin{center}
{\textbf{3.  Exact solutions and their application.}} \end{center}

In this section, we apply the operators of conditional symmetry for constructing exact solutions of the
relevant RD systems, demonstrate their  application for solving boundary value problems
and present  some biological interpretation.

It is well-known that any $Q$-conditional symmetry operator 
of the given two-dimensional PDE (system of PDEs)
guarantees its reduction to  an ODE (system of ODEs) by construction of the corresponding
local substitution, which is called ansatz. Usually  the ODE (system of ODEs) obtained can be
integrated (at least partly) and   exact solutions of the initial PDE (system of PDEs)
are constructed.
Nevertheless, this procedure  contains new difficulties,
 one can successfully realize one for the operators arising in table 1.
 We present the final result in the form of table 2.

 {\bf Table 2.  Ans\"atze and reduced systems of ODEs  for  RD systems of the form  (\ref{1})}

 \begin{center}
\begin{tabular}{|c|c|c|
}
\hline
no &   Ans\"atze  & Systems of ODEs    \\
\hline
 && \\
1. & $U=(\vp(x))^{1\over{k+1}}$&  $\vp_{xx}=-(k+1)f(\vp)$ \\
  &
 $V=(p^2 t+\psi(x))^2$&$\psi_{xx}=-{1\over2}g(\vp)+(\lambda+p)\psi $ \\
\hline
 && \\

2. &  $U=(\lbd_1(k+1) t+\vp(x))^{1\over{k+1}} $ &$\vp_{xx}=-(k+1)f(\vp-\al \psi)$  \\
  &$
V=(\lbd_2(l+1) t+\psi(x))^{1\over{l+1}}$ & $\psi_{xx}=-(l+1)g(\vp-\al \psi)$\\
\hline
 && \\

3. &  $U=(p^2 t+\vp(x))^2$&  $\vp_{xx}=(p+\lbd)\vp-{1\over2}f(\vp-\psi) $ \\
  &$
V=(p^2 t+\psi(x))^2$ &   $\psi_{xx}=(p+\lbd)\psi-{1\over2}g(\vp-\psi)$ \\
\hline
 && \\

4.&
$U=(\lbd_1(k+1) t+\vp(x))^{1\over{k+1}}$ & $\vp_{xx}=-(k+1)f\lf(\exp(\vp-\frac{\lbd_1(k+1)}{\lbd_2(l+1)}\psi)\rg)$ \\
 & $
V=(\exp(\lbd_2(l+1) t+\psi(x))+\lbd_3)^{1\over{l+1}}$ &
$\psi_{xx}+\psi_x^2=-(l+1)g\lf(\exp(\vp-\frac{\lbd_1(k+1)}{\lbd_2(l+1)}\psi)\rg)$
 \\
\hline
 && \\

5.&
$U=(\vp(x)\exp(\lbd_2(k+1) t)+\lbd_1)^{1\over{k+1}}$  & $\vp_{xx}=-(k+1)\vp f\lf(\vp\psi^{-\frac{\lbd_2(k+1)}{\lbd_4(l+1)}}\rg)$ \\
 & $
V=(\psi(x)\exp(\lbd_4(l+1) t)+\lbd_3)^{1\over{l+1}}$ & $\psi_{xx}=-(l+1)\psi g\lf(\vp\psi^{-\frac{\lbd_2(k+1)}{\lbd_4(l+1)}}\rg) $  \\

\hline
\end{tabular}
\end{center}

{\textbf{ Remark 2.}} The numbers 1,\dots,5 in table 2 correspond to the RD systems
arising in table 1 with the same numbers.

Now one sees that the reduced  systems of ODEs are nonlinear and it is quite implausible
that those are integrable for arbitrary smooth functions $f $ and $g $.
However, these systems can be integrated if the functions $f$ and $g$ are correctly specified.
For example, the ODE system arising in case 5  takes the form
\be\label{22}\ba{l}\vp_{xx}=\al_1\vp+\beta_1\psi,\\
\psi_{xx}=\al_2\vp+\beta_2\psi,\ea\ee if one sets
$g=-\frac{1}{l+1}(\al_2{\vp\over\psi}+\beta_2)
,\
f=-\frac{1}{k+1}(\al_1+\beta_1{\psi\over\vp}),
\,  r=-\lbd_2(k+1)=-\lbd_4(l+1),$
where $ \al_j,\ \beta_j, \,
 j=1,2$  are arbitrary constants.
 Nevertheless the initial RD system
 \be\label{33}\ba{l} U_t=(U^k U_x)_x-\frac{1}{k+1}(r U+\al_1(U^{k+1}-\lbd_1)+\beta_1(V^{l+1}-\lbd_3)
 -\lbd_1 r U^{-k}),\\
V_t=(V^l V_x)_x-\frac{1}{l+1}(r V+ \al_2(U^{k+1}-\lbd_1)+\beta_2(V^{l+1}-\lbd_3)-\lbd_3 r
V^{-l})\ea\ee
is still non-linear, the reducing system  (\ref{22}) is linear and its general solution can be
constructed in explicit form. Assuming $\beta_1\not=0$ (otherwise should be $\al_2\not=0$
and  one will start from the second equation of (\ref{22})) the function $\psi$ can be expressed
from the first  equation of (\ref{22}) so that the second equation takes the form 
\be\label{23}\vp_{xxxx}-(\al_1+\beta_2)\vp_{xx}+(\al_1\beta_2-\al_2\beta_1)\vp=0.\ee
The general solution of  the fourth order
ODE  (\ref{23}) essentially depends on the roots
of the algebraic equation
\be\label{24}s^4-(\al_1+\beta_2)s^2+(\al_1\beta_2-\al_2\beta_1)=0.\ee
Generally speaking, 9 different forms of the function $\vp$ can be
obtained. To avoid cumbersome computations, we consider only the case when
 (\ref{24}) possesses four different complex roots with the zero real parts, i.e.:
\be\label{24*}\ba{l}
s_{1,2}=\pm
i\sqrt{\frac{-(\al_1+\beta_2+\sqrt{(\al_1-\beta_2)^2+4\al_2\beta_1})}{2}},\\
s_{3,4}=\pm
i\sqrt{\frac{-(\al_1+\beta_2-\sqrt{(\al_1-\beta_2)^2+4\al_2\beta_1})}{2}}, \, i^2=-1.\ea\ee
It occurs under  the following restrictions on the coefficient of   (\ref{23}):
\be\label{24**} (\al_1-
\beta_2)^2>-4\al_2\beta_1,\quad
\al_1+\beta_2<-\sqrt{(\al_1-\beta_2)^2+4\al_2\beta_1}.\ee
Hence, we obtain the general solution of (\ref{22}):

\be\label{28}\ba{l}\vp=A_1 \cos(|s_1| x)+A_2
\sin(|s_1| x)+A_3 \cos(|s_3| x)+A_4 \sin(|s_3| x), \\
\psi=-\frac{1}{\beta_1}\biggr((\al_1+|s_1|^2)(A_1\cos(|s_1|
x)+A_2\sin(|s_1| x))+\\ \qquad+(\al_1+|s_3|^2)(A_3\cos(|s_3|
x)+A_4\sin(|s_3| x))\biggr),\ea\ee where $A_j
\, j=1,2,3,4 $  are arbitrary constants.
 Substituting (\ref{28}) into ansatz
\be\label{25}\ba{l}U=(\vp(x)\exp(\lbd_2(k+1) t)+\lbd_1)^{1\over{k+1}},\\
V=(\psi(x)\exp(\lbd_4(l+1) t)+\lbd_3)^{1\over{l+1}},\ea\ee arising
in case 5 of table 2, we arrive at the fourth-parametrical family of
exact solutions of the nonlinear RD system (\ref{33}) with the
coefficients satisfying restrictions   (\ref{24**}). It should be
noted that all solutions of the form (\ref{28})--(\ref{25})  are
periodic on the space variable $x$. According to J.D. Murray \cite{mur2},
  the periodic (in space) solutions   can mathematically express the Turing instability,
which leads to the spatially inhomogeneous  structures and forms. Moreover, we can show that
 solutions  of the form (\ref{28})--(\ref{25}) with the correctly specified constants $A_j, j=1 \dots 4$
 satisfy the zero Neumann
conditions (zero flux boundary conditions), which naturally arise in mathematical models with
the Turing instability. In fact, consider the space interval $[0,a], \, a>0$ where the cells with the
density $U$  and chemicals with the
density $V$ are interacting. Assuming that their interaction is described by the RD system  (\ref{33}),
the zero flux boundary conditions on $[0,a]$ lead to the requirements
\be\label{28**}\ba{l} U_x|_{x=0}=0,\quad  V_x|_{x=0}=0,\\
U_x|_{x=a}=0, \quad  V_x|_{x=a}=0.\ea \ee Using ansatz (\ref{25}),
one obtains formulas
\be\label{29}\ba{l}U_x=\frac{1}{k+1}\vp_x(\vp\exp(-rt)+\lbd_1)^{-\frac{k}{k+1}},\\
V_x=\frac{1}{l+1}\psi_x(\psi\exp(-rt)+\lbd_3)^{-\frac{l}{l+1}}.\ea\ee
Hence, taking into account  solution (\ref{28}) and  conditions
(\ref{28**}) with $x=0$, we arrive at two linear algebraic equations 
\[ \vp_x|_{x=0}=
A_2|s_1|+A_4|s_3|=0,\]
\[\psi_x|_{x=0}=A_2|s_1|(\al_1+|s_1|^2)+A_4|s_3|(\al_1+|s_3|^2)=0 \]
 to find  $A_2$ and
$A_4.$ Analyzing these equations   one easily finds that  the
solution $A_2=A_4=0$ only is possible. Finally, using conditions
(\ref{28**}) with $x=a$, we obtain
\be\label{29*} \ba{l} A_1|s_1|\sin(|s_1|a)+A_3|s_3|\sin(|s_3|a)=0,\\
A_1|s_1|(\al_1+|s_1|^2)\sin(|s_1|
a)+A_3|s_3|(\al_1+|s_3|^2)\sin(|s_3| a)=0\ea \ee
 to find $A_1$ and
$A_3$. Analyzing (\ref{29*}), we deduce that the space interval
$[0,a], \, a>0$ cannot be arbitrary otherwise $\vp=\psi=0$. There
are three countable sets of values for $a$: {\it(i)} $a=\frac{\pi
j_1}{|s_1|}$, \, {\it(ii)} $a=\frac{\pi j_1}{|s_3|}$ and {\it(iii)}
$a=\frac{\pi\sqrt{j_1^2+j_2^2}}{\sqrt{-(\al_1+\beta_2)}},\ (j_1, j_2)\subset
\mathbb{N}^2,$ for which  condition (\ref{29*}) is satisfied  by the
functions
\[ \vp=A_1 \cos(|s_1| x), \quad
\psi=-\frac{1}{\beta_1}A_1(\al_1+|s_1|^2)\cos(|s_1|x),\]
\[ \vp=A_3 \cos(|s_3| x),\quad
\psi=-\frac{1}{\beta_1}A_3(\al_1+|s_3|^2)\cos(|s_3|x),\] and
\[\ba{l}  \vp=A_1\cos(|s_1|x)+A_3\cos(\frac{j_1}{j_2}|s_1|x),\\
\psi=-\frac{1}{\beta_1}\biggr(A_1(\al_1+|s_1|^2)\cos(|s_1|x)+A_3(\al_1+\frac{j_1^2}{j_2^2}|s_1|^2)\cos(\frac{j_1}{j_2}|s_1|x)\biggr),\ea\]
respectively. Moreover, the additional restriction
$\al_2=\frac{(\al_1 j_2^2-\beta_2 j_1^2)(\beta_2
j_2^2-\al_1j_1^2)}{(j_1^2+j_2^2)^2\beta_1}$ is obtained in the case
{\it(iii)}, which allows to simplify the expression for $|s_1|$ and
$|s_3|$: \be\label{29**}
|s_1|=\frac{j_2\sqrt{-(\al_1+\beta_2)}}{\sqrt{j_1^2+j_2^2}}, \quad
|s_3|=\frac{j_1}{j_2}|s_1|. \ee

Thus,  we can now formulate the theorem.

\begin{theo}
The nonlinear RD system (\ref{33}) with the coefficients satisfying restrictions   (\ref{24**})
possesses the periodic (in space) solutions\\

\be\label{34}\ba{l}{\it(i)}\ U=\biggr(A_1\cos(|s_1|x)\exp(-r t)+\lbd_1\biggr)^{1\over{k+1}},\\
\quad \
V=\Biggr(-(A_1(\al_1+|s_1|^2)\cos(|s_1|x))\frac{\exp(-rt)}{\beta_1}+\lbd_3\Biggr)^{1\over{l+1}},\ea\ee

 satisfying the zero boundary  conditions (\ref{28**}) on the interval $\lf[0,\frac{\pi
j_1}{|s_1|}\rg],\ j_1\in\mathbb{N}$;\\
 \be\label{35}\ba{l}{\it(ii)}\ U=\biggr(A_3\cos(|s_3|x)\exp(-r t)+\lbd_1\biggr)^{1\over{k+1}},\\
\quad \ \
V=\Biggr(-(A_3(\al_1+|s_3|^2)\cos(|s_3|x))\frac{\exp(-rt)}{\beta_1}+\lbd_3\Biggr)^{1\over{l+1}},\ea\ee
 satisfying these conditions  on the interval $\lf[0,\frac{\pi
j_1}{|s_3|}\rg],\ j_1\in\mathbb{N}$;\\
\be\label{32}\ba{l}{\it(iii)}\ U=\biggr((A_1\cos(|s_1|x)+A_3\cos(\frac{j_1}{j_2}|s_1|x))\exp(-r t)+
\lbd_1\biggr)^{1\over{k+1}},\\
V=\lf(-\biggr(A_1(\al_1+|s_1|^2)\cos(|s_1|x)+A_3(\al_1+\frac{j_1^2}{j_2^2}|s_1|^2)
\cos(\frac{j_1}{j_2}|s_1|x)\biggr)
\frac{\exp(-rt)}{\beta_1}+\lbd_3\rg)^{1\over{l+1}},\ea\ee  satisfying these conditions on
the interval
$\lf[0,\frac{\pi\sqrt{j_1^2+j_2^2}}{\sqrt{-(\al_1+\beta_2)}}\rg],\ (j_1,
j_2)\subset \mathbb{N}^2.$

In the cases {\it(i)}and  {\it(ii)}, the values of  $|s_1|$ and $ |s_3|$ are determined  by formulae
(\ref{24*}), in the case  {\it(iii)} the value of  $|s_1|$ is given in (\ref{29**}) and
$\al_2=\frac{(\al_1 j_2^2-\beta_2 j_1^2)(\beta_2
j_2^2-\al_1j_1^2)}{(j_1^2+j_2^2)^2\beta_1}.$

\end{theo}

{\textbf{ Remark 3.}} Since system (\ref{33}) and boundary conditions (\ref{28**})
are  invariant under the space translation transformations
 $x \to x-x_0,$  where   $x_0$ is an arbitrary parameter, one can generalize this theorem
  on the intervals $[-a, 0]$ and  $[-a, a]$.

\begin{figure}[t]\label{Fig-1}
\begin{minipage}[t]{8cm}
\centerline{\includegraphics[width=8cm]{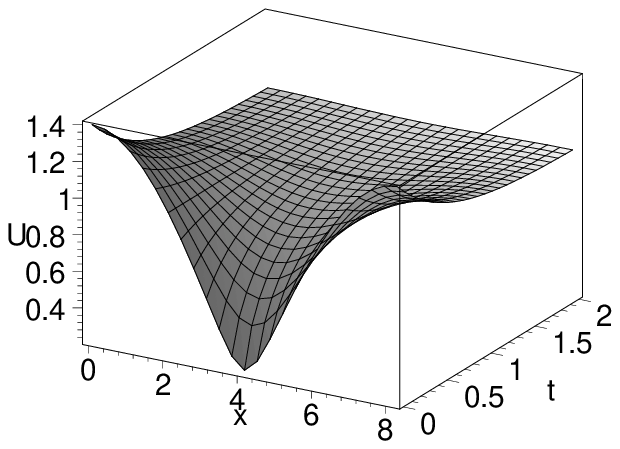}}
\end{minipage}
\hfill
\begin{minipage}[t]{8cm}
\centerline{\includegraphics[width=8cm]{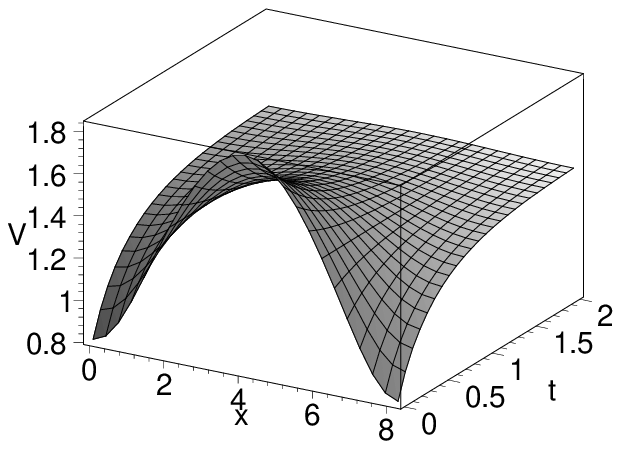}}
\end{minipage}
\center\caption{ Exact solution (\ref{34}) with $r=2,\ A_1=0.95,\
s_1=\sqrt{2-\sqrt{2}},\ \lbd_1=1,\ \lbd_3=2$}
\end{figure}

One sees that each solution of the forms (\ref{34})--(\ref{32})
tends  to the steady state point $(U,V)=((\lambda_1)^{1\over{k+1}},
(\lambda_3)^{1\over{l+1}})$ if the time $t \to +\infty$ and   $r>0$.
This point can be stable or unstable depending on the coefficients
of system (\ref{33}). An example of solution (\ref{34})
of the RD
system\\
\be\label{36}\ba{l}
\medskip
U_t=(UU_x)_x-\frac{r}{2}U+ U^2+\frac{1}{2}V^2+\frac{r-4U}{2U},\\
\medskip
V_t=(VV_x)_x-\frac{r}{2}V+U^2+V^2+\frac{r-3V}{V} \ea\ee 
with with $r=2$ is presented
on Fig.1. This solution  arrives at the  steady state point
$(U,V)=(1, \sqrt{2})$ very fast.

Solutions (\ref{34})--(\ref{32}) with $r<0$ are growing to infinity
with time providing $k+1>0$ and $l+1>0$. On the other hand, the
spatially inhomogeneous  structures should be formed for a finite
time (the development of those structures in an organism should be
finished  during the organism life). Taking this into account, one
notes that solutions (\ref{34})--(\ref{32}) with $r<0$ may describe
the development of spatially inhomogeneous  structures for the
finite time $t_{max}$. The relevant example of solution (\ref{34})
of the RD system (\ref{36}) with $r=-2$ is presented on Fig.2.

\begin{figure}[t]\label{Fig-2}
\begin{minipage}[t]{8cm}
\centerline{\includegraphics[width=8cm]{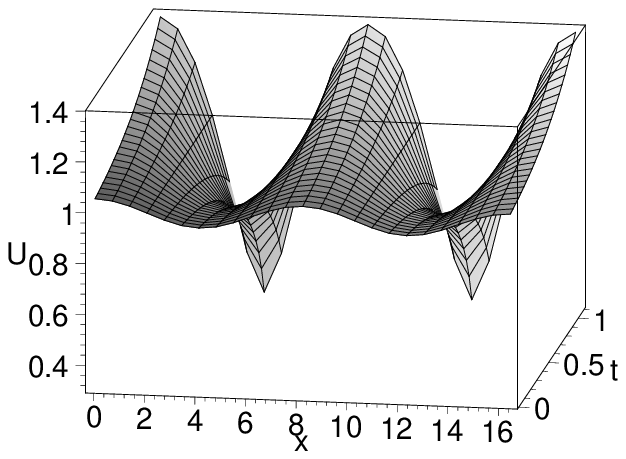}}
\end{minipage}
\hfill
\begin{minipage}[t]{8cm}
\centerline{\includegraphics[width=8cm]{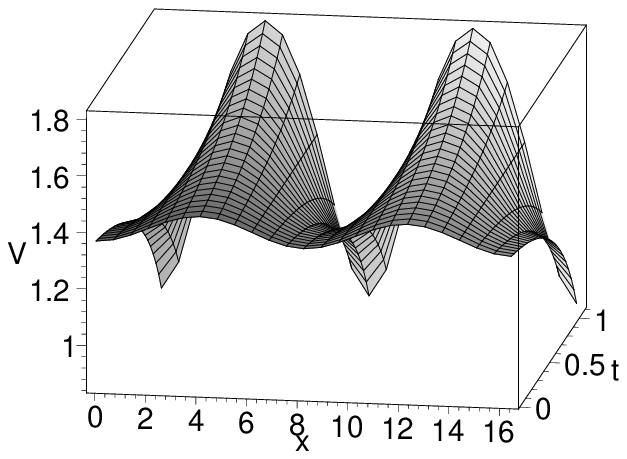}}
\end{minipage}
\center\caption{ Exact solution (\ref{34}) with $r=-2,\ A_1=0.1,\
s_1=\sqrt{2-\sqrt{2}},\ \lbd_1=1,\ \lbd_3=2$}
\end{figure}

 One sees that both components, $U$ and $V$ have practically  homogeneous
 distributions in space if  $t<< t_{max}$. However, those are essentially  inhomogeneous
  if $t>>0$. Moreover each of the components $U$ and $V$ dominates in the different regions.
 So, we may  say that  this solution  forms a striped structure. This kind of spatially inhomogeneous  structures
 is typical for mammalian coat (tigers, zebras).

\begin{center}
{\textbf{4. Conclusions.}}
\end{center}

In this paper, theorem 1  giving a complete description of
$Q$-conditional symmetries of the nonlinear RD system (\ref{1})
under some restrictions  is proved. We established that there is a wide range of
the nonlinear RD systems of five types, which are conditionally invariant with respect to the
operator  (\ref{10}) with the coefficients $\xi=0$ and $\eta^1_V=\eta^2_U=0$.
Moreover those systems don't coincide with the RD systems  admitting a non-trivial Lie
symmetry, which has been completely described in  \cite{ibrag-94, ch-king4}.
To our best knowledge, there are no any papers devoted to constructions
of conditional symmetries
 for  {\it systems of  evolution equations}, excepting the recent paper
 \cite{ch-se-03}. Note that only an example of the nonlinear  diffusion-convection system  was found
 in \cite{ch-se-03}, while here we present five classes of the nonlinear RD systems
 and each of them involves two arbitrary functions.
 The work is in progress to solve the problem without any restrictions on the coefficients
 of operator (\ref{10}).

The  $Q$-conditional symmetry operators have been applied for obtaining exact solutions
of the relevant RD systems. In fact, we reduced the problem of solving
 PDE systems to one  for the  relevant ODE systems. 
 These ODE systems were further solved therefore several exact solutions
 in explicit form have been constructed for the RD systems (\ref{33}).
  Properties of the solutions obtained were investigated and a possible biological
 interpretation  was suggested.

\medskip

\centerline {\bf Acknowledgments}

\medskip

The authors are grateful to the unknown board member and the unknown
referee for the  useful comments.

\end{document}